\documentclass{appolb}
\usepackage{graphicx}

\usepackage{amsmath}
\usepackage{amssymb}
\usepackage{amsfonts}
\usepackage{graphicx}
\usepackage{pstricks}
\usepackage{tikz}
\usepackage{mathtools}

\newcommand{\be}{\begin{equation}}
\newcommand{\ee}{\end{equation}}
\newcommand{\bqr}{\begin{eqnarray}}
\newcommand{\eqr}{\end{eqnarray}}

\begin{document}

\title{Linear response theory for Gogny interaction}

\author{D. Davesne$^{a,b}$, J. Navarro$^{c}$, A. Pastore$^{d}$
\address{$^{a}$ Universit\'e de Lyon, Universit\'e Lyon 1, 43 Bd. du 11 Novembre 1918, F-69622 Villeurbanne cedex, France\\
$^{b}$              CNRS-IN2P3, UMR 5822, Institut de Physique Nucl{\'e}aire de Lyon \\
$^{c}$ IFIC (CSIC-Universidad de Valencia), Apartado Postal 22085, E-46.071-Valencia, Spain \\
$^{d}$ Department of Physics, University of York, Heslington, York, Y010 5DD, United Kingdom}
}
\maketitle


\begin{abstract}
We present the formalism of the linear response theory in symmetric nuclear matter for a finite-range central interaction interaction including  zero-range spin-orbit and tensor  components.
\end{abstract}
  

\section{Introduction}

Symmetric nuclear matter (SNM) is a spin-saturated system composed of an equal amount of neutrons and protons with no finite-size effect. Although it may appear as an highly ideal system, the study of its excitation modes provides us with several useful guidelines on other systems as atomic nuclei and neutron stars (NS).

In Ref.~\cite{report}, we have presented in great detail the formalism of Linear Response (LR) theory for a Skyrme functional including both spin-orbit and tensor terms~\cite{les07}. By studying the response function of the system to an external probe, we have been able to identify the critical densities and momenta at which instabilities  occur in the system~\cite{pas12} and relate them to the appearance of finite-size instabilities in nuclei~\cite{hel13,Pas15T}.
We have  then provided in Refs.~\cite{pas13,bec17b} a simple method based on LR in SNM to avoid such instabilities directly at the level of the fitting procedure.

The response functions of SNM, or more generally of asymmetric nuclear matter~\cite{dav14}, have also direct application to study neutrino opacities in NS~\cite{iwa82,pas12b,pas14SK} which in turn have important effects on NS cooling~\cite{sht11}.

In the present article, we present an extension of the formalism of LR theory to the case of  finite-range interactions with spin-orbit and tensor terms. The basic formalism has been discussed in Ref.~\cite{mar05}, for the case of a central part of the Gogny interaction~\cite{dec80}. The authors of Ref.~\cite{dep16} have already discussed the LR of a Gogny interaction using continued fraction (CF) approximation~\cite{mar08}. The main inconvenient of such a method is that most of the integrations required to obtain the response functions need to be performed numerically via Monte-Carlo samplings. Moreover, in Ref.~\cite{dep16}, the spin-orbit term has been neglected, which is only valid for low-transfer momenta, see discussion in Ref.~\cite{pas12}.
The current formalism, being based on partial wave decomposition~\cite{dav15s}, offers us the opportunity of including all terms of the interaction. Compared to Ref.~\cite{dep16}, our formalism  requires more detailed analytical derivations of all matrix elements in the different channels, but it is numerically less expensive.

The article is organised as follows: in Sec.~\ref{sec:formalism} we briefly introduce the LR formalism using multipolar expansion, while in Sec.~\ref{sec:result}, we illustrate our results for the case of a Gogny interaction. We present our conclusions in Sec.~\ref{sec:con}.


\section{Formalism}\label{sec:formalism}

The response function of the system is obtained by integrating, over the momentum, the RPA propagator~\cite{report,gar92}.
The latter is the solution of the Bethe-Salpeter equation in each spin (S), spin-projection (M) and isospin (I) channel, for brevity called $\alpha\equiv\text{(S,M,I)}$ in this paper. It reads

\begin{eqnarray}\label{eq:be}
G^{(\alpha)}_{\text{RPA}}(q,\omega,\mathbf{k}_1)&=&G_{\text{HF}}(q,\omega,\mathbf{k}_1)\\
&+&G_{\text{HF}}(q,\omega,\mathbf{k}_1)\sum_{\alpha'}\int \frac{d^3 \mathbf{k}_2}{(2\pi)^3}V_{ph}^{(\alpha,\alpha')}(q,\mathbf{k}_1,\mathbf{k}_2)G^{(\alpha')}_{\text{RPA}}(q,\omega,\mathbf{k}_2)\;,\nonumber
\end{eqnarray}

\noindent where $q$ is the transferred momentum and $\mathbf{k}_1,\mathbf{k}_2$ is the momentum of the particle-hole pair and $\omega$ the transferred energy. $V_{ph}^{(\alpha,\alpha')}(q,\mathbf{k}_1, \mathbf{k}_2)$ represents the residual particle-hole interaction.
In the case of a Skyrme interaction~\cite{report,pas12,gar92,bec15}, Eq.~(\ref{eq:be}) can be solved analytically using a system of symbolic equations.
For the case of a finite-range interaction as Gogny, this is not possible. We have thus adopted the technique presented in Ref.~\cite{mar05} and performed a multipolar expansion as

\begin{eqnarray}
G_{\text{HF}}(q,\omega,\mathbf{k}_1)&=&\sum_l G^{\text{HF}}_l(q,\omega, k_1)Y_{l0}(\cos \theta_1)\\
V_{ph}^{(\alpha,\alpha')}(q,\mathbf{k}_1,\mathbf{k}_2)&=&\sum_{lm,l'm'}V^{(\alpha,\alpha')}_{lm;l'm'}(k_1,k_2)Y_{lm}(\hat{k}_1)Y^*_{l'm'}(\hat{k}_2)\\
G^{(\alpha)}_{\text{RPA}}(q,\omega,\mathbf{k}_1)&=&\sum_{lm}G^{(\alpha)}_{lm}(q,\omega,k_1)Y_{lm}(\hat{k}_1)
\end{eqnarray}

\noindent where $Y_{lm}$ is the usual spherical harmonic. We then obtain the multipolar expansion of Eq.~(\ref{eq:be}) as

\begin{eqnarray}\label{eq:multi}
G_{lm}^{(\alpha)}(k_1)=\delta_{m,0}G^{\text{HF}}_l(k_1)+\sum_{\alpha'}\sum_{l'm'}\int \frac{d k_2 k_2^2}{(2\pi)^3}M_{lm,l'm'}^{(\alpha,\alpha')}(k_1,k_2)G_{l'm'}^{(\alpha')}(k_2) \, ,
\end{eqnarray}

\noindent where, for simplicity, we have dropped the explicit dependence on $q$ and $\omega$. The matrix elements $M_{lm,l'm'}^{(\alpha,\alpha')}$ are defined as

\begin{eqnarray}
M_{lm,l'm'}^{(\alpha,\alpha')}(k_1,k_2) &=& \sum_{l_1l_2}\left[\frac{(2l_2+1)(2l_1+1)}{4\pi(2l+1)} \right]^{1/2}C^{l0}_{l_10l_20}C^{lm}_{l_2m,l_10}
\nonumber \\
&& \hspace{3cm} \times G^{HF}_{l_1}(k_1)V^{(\alpha,\alpha')}_{lm;l'm'}(k_1,k_2)
\end{eqnarray}

\noindent The notation used here simplifies to the one of Ref.~\cite{mar05} in the case of a central interaction for which $l=l'$ and $m=m'$.
Eq.~(\ref{eq:multi}) actually represents a system of coupled integral equations. To solve them, we discretise the integrals over a uniform mesh over the range $k\in[0,q+k_F]$ as done in Ref.~\cite{mar05} and then invert the matrix of the system. 
In principle, the size of the system is infinite, since a finite range interaction contains all multipoles. However, as discussed in Ref.~\cite{mar05}, the contribution of higher order multipoles becomes smaller and smaller. We have thus introduced a cut-off parameter $L_{Max}$ to limit the number of equations we have to solve.
In Sec.~(\ref{sec:result}), we discuss the convergence of our results with $L_{Max}$.

\section{Results}\label{sec:result}

To test the quality of our results, we start by considering the case of a Skyrme force with tensor term, namely T44~\cite{les07}. Since the Skyrme interaction is a simple combination of S and P waves~\cite{dav16ANNALS}, the partial wave expansion has a natural truncation. Notice that additional caution should be taken when considering an explicit tensor contribution : since the tensor term $t_e$ couples $S$ and $D$ waves, we have to take $L_{Max}=2$.
Beyond $L=2$ the other contributions are zero, due to the particular form of the Skyrme force.

\begin{figure}[h!]
\centerline{%
\includegraphics[width=0.55\textwidth,angle=-90]{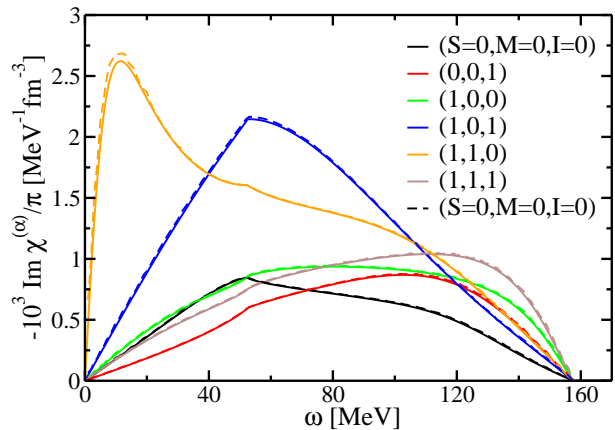}
}
\caption{(Colors online) Response functions $-\text{Im}\chi^{(\alpha)}/\pi $ in SNM calculated at saturation density and transferred moment $q=k_F$ for Skyrme T44 interaction. The small differences between solid and dashed lines are the consequence of the method (discretisation) used to solve Bethe-Salpeter equations. See text for details.}
\label{t44}
\end{figure}

In Fig.~\ref{t44}, we compare the response functions obtained with the method given in Eq.~(\ref{eq:multi}) (dashed lines) with those obtained by using the technique described in Ref.~\cite{report}. Both calculations are done at saturation density and transferred momentum  $q=k_F$.
The results stay on top of each other : the small deviations are only due to the discretisation on a finite grid of Eq.~(\ref{eq:multi}). This example clearly proves the validity of our method based on multipolar expansion and we will now apply it to the case of finite-range interactions.

\begin{figure}[h!]
\centerline{%
\includegraphics[width=0.45\textwidth,angle=-90]{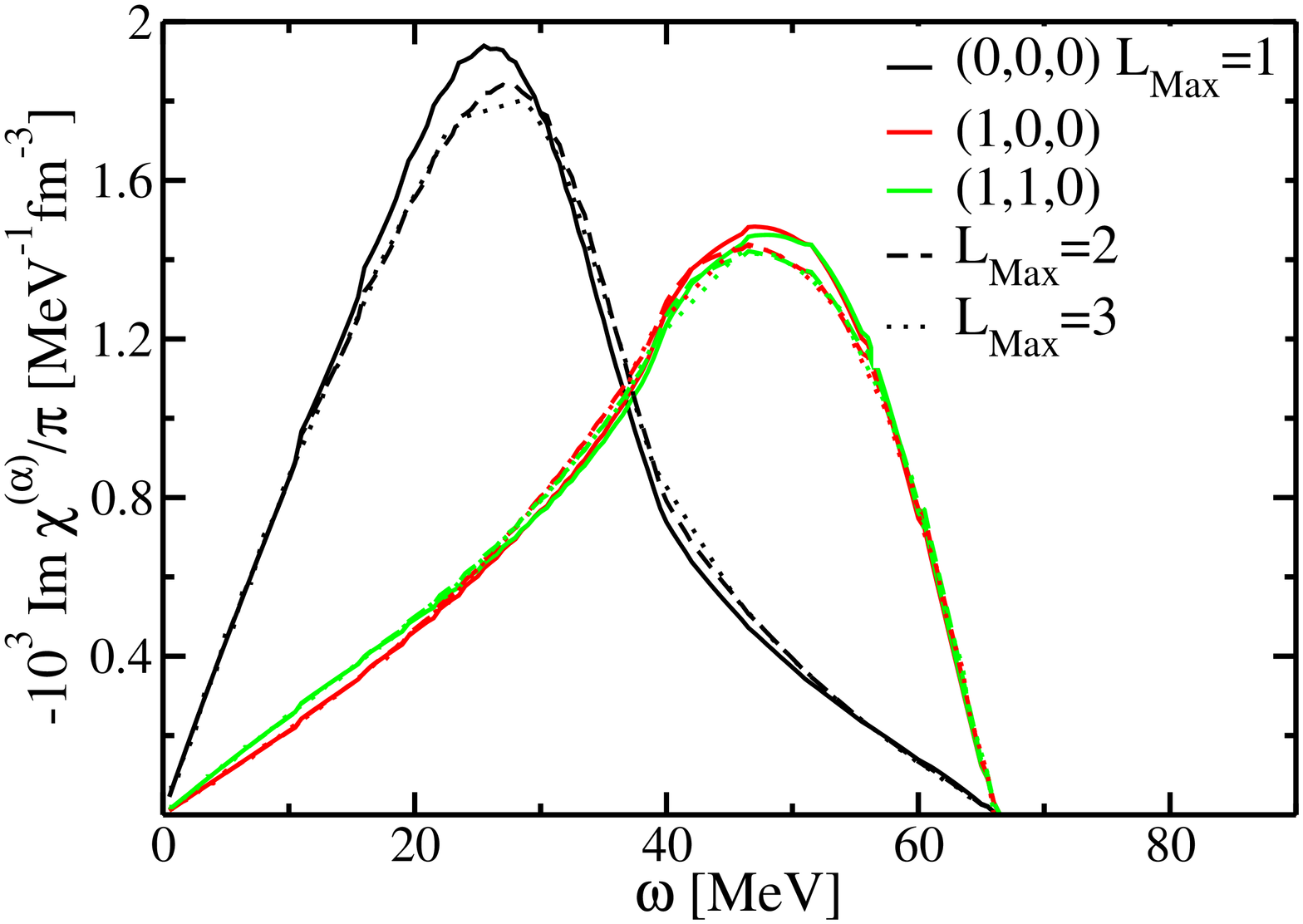}
\includegraphics[width=0.45\textwidth,angle=-90]{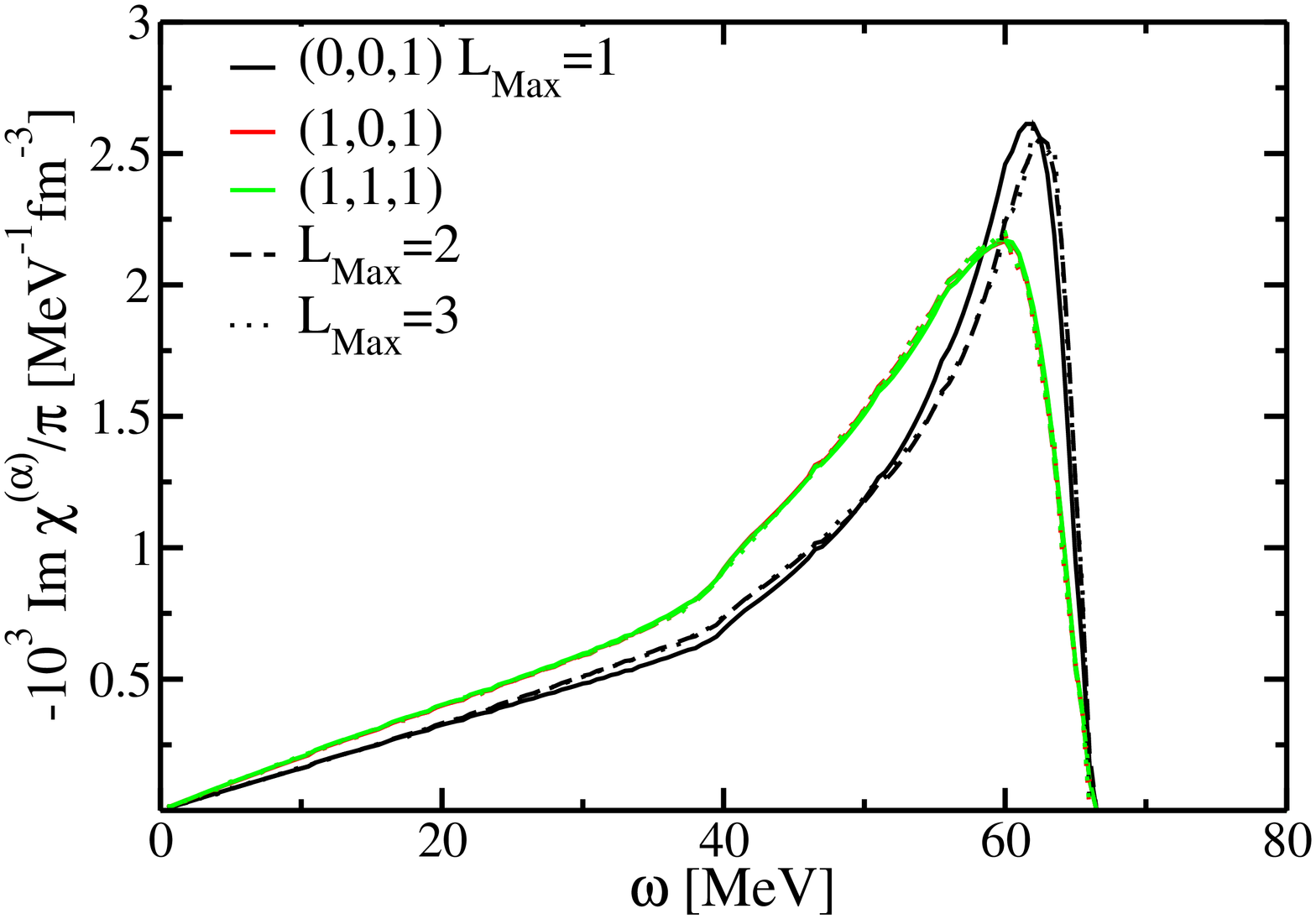}
}
\caption{(Colors online) Response function $-\text{Im}\chi^{(\alpha)}/\pi $ in SNM calculated at $\rho=0.16$ fm$^{-3}$ and transferred moment $q=0.5k_F$ for Gogny D1S interaction as a function of cut-off in angular momentum expansion $L_{Max}$. }
\label{d1s}
\end{figure}

\begin{figure}[h!]
\centerline{%
\includegraphics[width=0.45\textwidth,angle=-90]{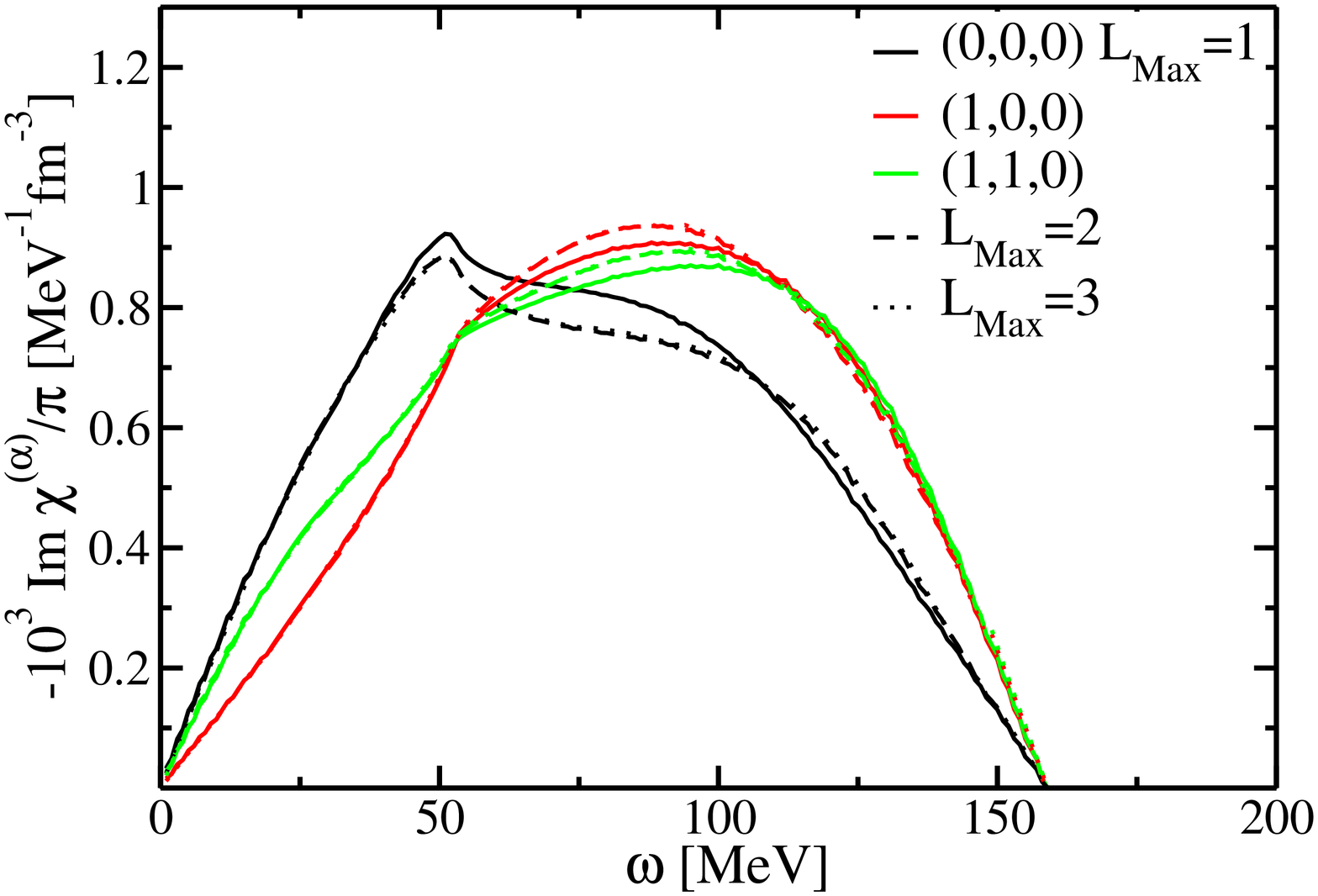}
\includegraphics[width=0.45\textwidth,angle=-90]{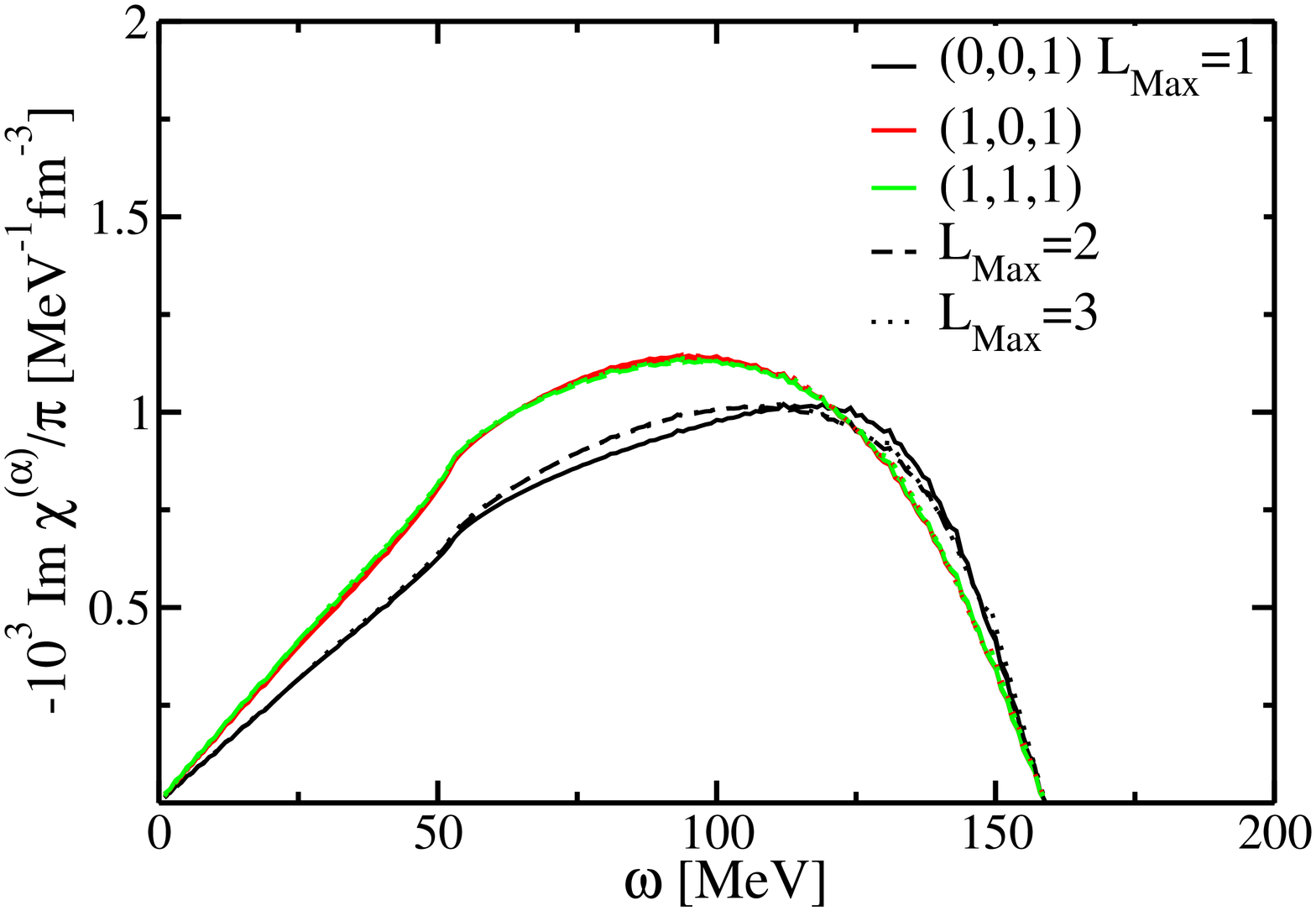}
}
\caption{(Colors online) Same as Fig.\ref{d1s}, but for $q=k_F$ .}
\label{d1s2}
\end{figure}

In Figs.~\ref{d1s}-\ref{d1s2}, we present the response functions in the different spin-isospin channels  for different values of the maximum angular momentum $L_{Max}$ in the case of Gogny D1S interaction~\cite{dec80}. The response functions are calculated at $\rho=0.16$ fm$^{-3}$ and at two different values of transferred momentum $q=0.1 k_F,q=k_F$. The Gogny D1S interaction is not equipped with a tensor term, thus the spin-orbit term only induces a splitting of the different projections of spin $M=0,\pm1$. Since the transferred momentum is still quite small, such a splitting is quite negligible as discussed in Ref.~\cite{dep16}.


\noindent We clearly observe that $L_{Max}=2$ is enough to obtain a reasonable description of the response function, confirming the results of Ref.~\cite{mar05}. 

\begin{figure}[h!]
\centerline{%
\includegraphics[width=0.55\textwidth,angle=-90]{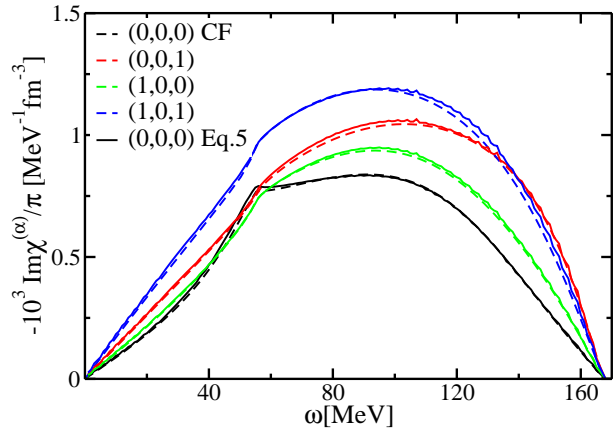}
}
\caption{(Colors online) Response function for the central term of the D1S Gogny interaction. The calculations are done at $\rho=0.174$ fm$^{-3}$ and $q=k_F$.  See text for details.}
\label{d1snorb}
\end{figure}

Finally we compare in Fig.~\ref{d1snorb} our calculations obtained using Eq.~(\ref{eq:be}) with a cut-off of $L_{Max}=2$ with the ones of Ref.~\cite{dep16}, based on Continued Fraction (CF) approximation~\cite{mar08}. Following Ref.~\cite{dep16}, the terms included in the calculations are the central part of the Gogny D1S interaction. The density is $\rho=0.174$ fm$^{-3}$ and the transferred momentum $q=k_F$. Without tensor or spin-orbit terms, there is no longer a splitting in the different $M$ projections of the total spin $S$. For such a reason we present only results with $M=0$. We can see that both methods are in excellent agreement, thus proving the validity of our results.

It is also interesting to compare the $S=0$ channels from Fig.~\ref{d1snorb} and Fig.~\ref{d1s2} since we can observe a non-negligible difference of the low-energy part of the isoscalar response function. To understand the role of spin-orbit in this channel, it is useful to observe the expression of the response function given in Ref.~\cite{pas12}: due to the spin-orbit term the interaction terms of the $S=1$ do contribute in the $S=0$ channel and the term mixing the two channels is typically proportional to the transferred momentum to the power of 4. As a consequence when the transferred momentum is small we can neglect such a term, but for large transferred momenta this term starts playing an important role.

\begin{figure}[h!]
\centerline{%
\includegraphics[width=0.55\textwidth,angle=-90]{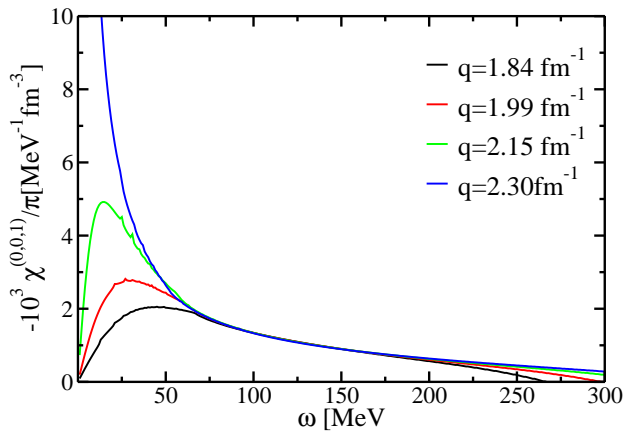}
}
\caption{(Colors online) Response functions for the Gogny D1S interaction the channel (0,0,1) and different values of the transferred momentum. See text for details.}
\label{pole}
\end{figure}

A very important aspect of the LR formalism is the detection of poles in the response functions. In Fig.~\ref{pole}, we considered the example of the Gogny D1S interaction and showed the evolution of the response function in the channel (0,0,1) for different values of transferred momentum.The calculations have been performed at $\rho=1.5\rho_0$, $\rho_0$  being the saturation density of the Gogny D1S interaction because it is known (see Ref.~\cite{dep16}) that at this particular value of density, D1S exhibits a pole at $q_c=2.53$ fm$^{-1}$. Effectively, we can observe the shape evolution of the response function when $q$ approaches the critical value $q_c$, showing the stability of our numerical approach and the ability of the formalism to detect instabilities. 

\section{Conclusions}\label{sec:con}

In this paper, we have presented the formalism of the LR theory for finite-range interactions using multipolar expansion of the Bethe-Salpeter equation. We have tested our formalism against the results of the LR formalism given in Ref.~\cite{report} for the case of Skyrme T44~\cite{les07}. In particular, thanks to the multipolar expansion, we clearly observe the tensor coupling between S and D waves.
As a further benchmark of the formalism, we have compared the result for the case of a Gogny interaction with Ref.~\cite{dep16}. In this case the results are in excellent agreement as well. The advantage of our technique (compared to Ref.~\cite{dep16}) is the explicit inclusion of spin-orbit term in the residual interaction. As explicitly shown, this may lead to non-negligible effects on the response function since it induces an extra coupling between the $S=0$ and $S=1$ channels.

Finally, we have explored how the current formalism describes the presence of poles in the response function. The goal is to continue in that direction so that we can include directly a test to detect instabilities in the fitting procedure itself. This can be important in the context of future Gogny-like parametrisation devoted to nuclear astrophysics by instance~\cite{vin18,mar18}.



\section*{Acknowledgments}


The work of J.N. has been supported by grant FIS2017-84038-C2-1-P, Mineco (Spain).
The work of A.P. is supported  by the UK Science and Technology Facilities Council under Grants No. ST/L005727 and ST/M006433. 

\bibliographystyle{polonica}
\bibliography{biblio}

\end{document}